# Three-period evolution in a photonic Floquet extended Su-Schrieffer-Heeger waveguide array

Changsen Li, [1] Yujie Zhou, [1] Xiumei Wang,[2*] and Xingping Zhou, [3*]

[1] *College of Integrated Circuit Science and Engineering, Nanjing University of Posts and Telecommunications, Nanjing 210003, China*

[2] *College of Electronic and Optical Engineering, Nanjing University of Posts and Telecommunications, Nanjing 210003, China*

[3] *Institute of Quantum Information and Technology, Nanjing University of Posts and Telecommunications, Nanjing 210003, China*

*\*Author to whom any correspondence should be addressed.*

*\*wxm@njupt.edu.cn*

*\*zxp@njupt.edu.cn*

Periodic driving can induce the emergence of topological $\pi$ modes, and their superposition with zero modes leads to two-period dynamics. Introducing long-range couplings enables the realization of larger topological winding numbers, which correspond to multiple pairs of degenerate edge states under open boundary conditions. In this work, we construct a Floquet extended Su-Schrieffer-Heeger (SSH) model by introducing a two-step periodic driving and next-nearest-neighbor coupling into the static SSH chain simultaneously. Remarkably, we identify anomalous edge states with quasienergies $\pm\pi/3T$ and $\pm 2\pi/3T$. In order to reveal the dynamical features of these anomalous edge states, we elaborately adjust the optical parameters and ultimately achieve a successful mapping of the model onto a photonic waveguide array. Subsequently, through numerical simulation of the wave equation, we observe the unique behavior of three-period evolution. Our work may serve as a reference for realizing period-multiplied dynamics, and the anomalous edge states discussed here might also find applications in quantum computation.

# I. INTRODUCTION

In the past decades, significant progress has been made in the investigation of Floquet systems both theoretically [1-7] and experimentally [8-14]. Owing to their excellent controllability, Floquet helical waveguide systems have emerged as a paradigmatic platform of novel photonic topological states and phenomena [15-23]. Floquet systems have time-periodic Hamiltonians satisfying $H(t+T) = H(t)$, where $T$ is the driving period. One important feature of them is the periodicity of their quasienergy spectrum along the energy axis. A static two-band system typically hosts only a single energy gap. However, due to the periodic nature of the quasienergy spectrum, a two-band Floquet system can accommodate two distinct gaps, i.e., the 0-gap and the $\pi$-gap [2, 3, 24, 50]. This necessitates the use of a pair of topological invariants to characterize the topological phases of Floquet systems. One of the most well-known examples is the two-dimensional anomalous Floquet topological insulator [2, 19, 25], whose bulk bands have zero Chern number but still support chiral edge states. To fully capture the topological phases of such models, researchers introduced new topological invariants associated with both the 0-gap and the $\pi$-gap [2].

Apart from differences in topological characterization, Floquet systems also exhibit a variety of interesting phenomena, such as dynamical localization [26-30], stabilization [31-36] and two-period evolution [37, 38]. The two-period evolution arises from the superposition of zero and $\pi$ modes. Recently, a class of Floquet topological edge states with quasienergy $\pm\pi/2T$, referred to as $\pm\pi/2$ modes, has been revealed [39-41]. The superposition of zero, $\pm\pi/2$ and $\pi$ modes can lead to four-period evolution. So far, the $\pm\pi/2$ modes have been observed in acoustic system [42] and photonic waveguide lattice [43]. We are interested in the evolution characteristics when the numbers of zero and $\pi$ modes are unequal. Previous works have shown that long-range couplings (LRCs) can induce multiple pairs of edge states [44-47]. Motivated by this, we construct a Floquet extended Su-Schrieffer-Heeger (SSH) model by introducing both periodic driving and next-nearest-neighbor (NNN) coupling into the one-dimensional static SSH chain. In this new model, we observe phases in which the numbers of zero and $\pi$ modes are different, as expected. In addition, we find anomalous edge states with quasienergies $\pm\pi/3T$ and $\pm 2\pi/3T$, which we refer to as the $\pm\pi/3$ and $\pm 2\pi/3$ modes. The existence of these states is expected to give rise to three-period evolutions.

The structure of this paper is as follows. In Sec. II, we introduce the Floquet extended SSH model and characterize its phases using a pair of winding numbers $(W_0, W_\pi)$. From the winding number phase diagram, we identify phases where the numbers of zero and $\pi$ modes differ. In Sec. III, we perform a parameter-space scan under open boundary conditions (OBC) and plot quasienergy spectra. Anomalous edge states are observed in the spectra, including those with quasienergies $\pm\pi/3T$ and $\pm 2\pi/3T$. In Sec. IV, we investigate the dynamical evolution properties of the system. The evolution results under the tight-binding approximation indicate that the superposition of $\pm\pi/3$ ($\pm 2\pi/3$) and $\pi$ (zero) modes can give rise to the three-period evolution. With carefully adjusted optical parameters, we map the Floquet extended SSH model onto a photonic waveguide array and successfully observe three-period evolution in the simulation. Considering the initial purpose of constructing the model, we also briefly analyze the two-period evolution when the numbers of zero and $\pi$ modes are unequal. In Sec. V, we provide a summary of our work and discuss its relevance to future research and possible applications.

## II. MODEL AND TOPOLOGICAL PHASE CHARACTERIZATION

### A. Model and topological invariants

We start by introducing the NNN coupling and a Floquet driving protocol into the traditional SSH chiral chain [48, 49]. As shown in Fig. 1, the model consists of two sublattices A and B in each unit cell. The Floquet driving protocol involves two steps of equal duration, each lasting $T/2$, where $T$ is the driving period. In the subsequent studies, we set $T=2$ for consistency in the analysis. We arrange the model in a bilayer structure to better illustrate the NNN coupling. The time-dependent Hamiltonian of the lattice is given by

$$H(t) = \begin{cases} H_1, & t \in [\ell T,\ \ell T+T/2) \\ H_2, & t \in [\ell T+T/2,\ \ell T+T) \end{cases}, \quad (1)$$

where

$$H_1 = u_0 \sum_{n=1}^{N} |n,\ B\rangle\langle n,\ A| + H.c.,$$
$$H_2 = u_1 \sum_{n=1}^{N-1} |n,\ B\rangle\langle n+1,\ A| + u_2 \sum_{n=1}^{N-2} |n,\ B\rangle\langle n+2,\ A| + H.c.. \quad (2)$$

Here, $\ell \in \mathbb{Z}$, $n \in \mathbb{Z}$ is the unit cell index, and $N \in \mathbb{Z}$ is the number of unit cells. The intracell coupling parameter $u_0$ is set to $\pi/T$, i.e., $\pi/2$. The $u_1$ and $u_2$ are the nearest-neighbor (NN) and NNN coupling parameters, respectively.

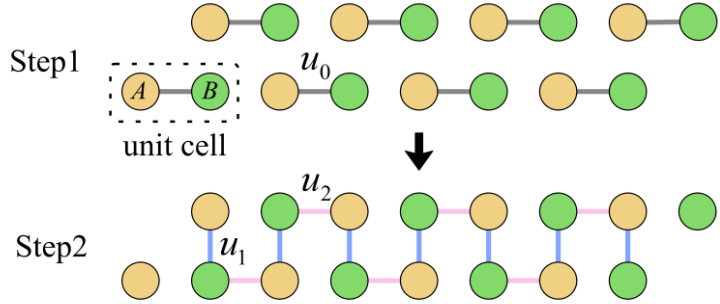

**FIG. 1.** Schematic illustration of Floquet extended SSH model. Each of the unit cells contains two sublattices A and B. In the first step, only the intracell couplings $u_0$ (gray) are active. In the second step, only the nearest-neighbor couplings $u_1$ (light blue) and next-nearest-neighbor couplings $u_2$ (pink) are active. Each step in the two-step driving protocol lasts for half of a full period.

A Floquet system is defined by a unitary evolution operator

$$U_T = \text{T}\exp\left(-i\int_0^T H(t)dt\right), \tag{3}$$

over one driving period, where $\text{T}$ is the time-ordering operator. In our model, only intracell couplings $u_0$ are present during the first half of the driving period, while only NN couplings $u_1$ and NNN couplings $u_2$ are present during the second half. Therefore, the Floquet operator describing the evolution of our model is given by $U_T = e^{-iH_2}e^{-iH_1}$. By solving the Floquet eigenvalue equation $U_T|\psi\rangle = e^{-i\varepsilon T}|\psi\rangle$, the quasienergy $\varepsilon$ can be obtained. Under periodic boundary conditions, the Floquet operator $U_T$ can also be expressed in momentum representation as $U_T = \sum_k U_T(k)|k\rangle\langle k|$, where

$$U_T(k) = e^{-iH_2(k)}e^{-iH_1(k)} = e^{-i\left[g_x(k)\sigma_x + g_y(k)\sigma_y\right]}e^{-iu_0\sigma_x}, \tag{4}$$

$$\begin{aligned}g_x(k) &= u_1\cos(k) + u_2\cos(2k),\\ g_y(k) &= u_1\sin(k) + u_2\sin(2k),\end{aligned} \tag{5}$$

and $k \in [-\pi, \pi)$ is restricted to the first Brillouin zone. The $\sigma_x$ and $\sigma_y$ are Pauli matrices.

Although both $H_1(k)$ and $H_2(k)$ possess chiral symmetry (CS) individually, the Floquet evolution operator $U_T(k)$ does not satisfy the CS condition $\Gamma U_T(k)\Gamma^{-1} = U^{-1}_T(k)$ due to the non-commutativity of $H_1(k)$ and $H_2(k)$. The $\Gamma$ is a unitary, Hermitian, and local (within a unit cell) operator. Similarly, the effective Hamiltonian $H_{\text{eff}}$ defined by

$$H_{\text{eff}}(k) = \frac{i}{T}\ln U_T(k), \tag{6}$$

does not possess CS either [50]. Each one-dimensional Floquet topological phase with CS is characterized by a pair of integer winding numbers $(W_0, W_\pi)$, which is defined in two symmetric time frames of the system's Floquet operator [51-53]. We can get the Floquet operators in these symmetric time frames as

$$U_1'(k) = e^{-i\frac{1}{2}H_1(k)} e^{-iH_2(k)} e^{-i\frac{1}{2}H_1(k)},$$
$$U_2''(k) = e^{-i\frac{1}{2}H_2(k)} e^{-iH_1(k)} e^{-i\frac{1}{2}H_2(k)}, \quad (7)$$

by shifting the starting time of the evolution forward or backward one quarter of the driving period. Thus, the effective Hamiltonians $H_{\text{eff}}'$ and $H_{\text{eff}}''$ defined via Eq. (6) have CS and can be assigned topological invariants $v'$ and $v''$. In the canonical basis, $H_{\text{eff}}$ can be written in a block off-diagonal form [50]:

$$H_{\text{eff}}(k) = \begin{pmatrix} 0 & h(k) \\ h^\dagger(k) & 0 \end{pmatrix}. \quad (8)$$

Based on $h(k)$, topological invariants $v'$ and $v''$ can be obtained from

$$v[h(k)] = \frac{1}{2\pi i} \int_{-\pi}^{\pi} dk \frac{d}{dk} \ln[h(k)]. \quad (9)$$

Finally, by combining $v'$ and $v''$, a pair of winding numbers $W_0$ and $W_\pi$ can be defined to characterize the topological phases of the Floquet extended SSH model:

$$W_0 = \frac{v' + v''}{2}, \quad W_\pi = \frac{v' - v''}{2}. \quad (10)$$

**B. Characteristics of phase distribution**

As shown in Fig. 2(a), we compute winding numbers $W_0$ and $W_\pi$ for our Floquet extended SSH model based on Eqs. (9) and (10). The range of parameters is $(u_1, u_2) \in [-6u_0, 6u_0] \times [-6u_0, 6u_0]$. We observe that both types of winding number distributions exhibit symmetry with respect to the $u_2$ axis. It indicates that the sign of the NN coupling parameter $u_1$ does not affect the topological outcome. Moreover, it is evident that the distributions of $W_0$ and $W_\pi$ are symmetric with respect to the $u_1$ axis, but with opposite signs. Therefore, if the topological invariants at $(u_1, u_2)$ are $(n, m)$, the invariants at $(-u_1, u_2)$ are $(n, m)$, at $(u_1, -u_2)$ are $(-m, -n)$ and at $(-u_1, -u_2)$ are likewise $(-m, -n)$. According to the bulk-boundary correspondence [51,52], the absolute value of the winding number corresponds to the number of degenerate edge-state pairs at the associated quasienergy under OBC, i.e.,

$$n_0 = |W_0|, \quad n_\pi = |W_\pi|. \tag{11}$$

Thus, when the coupling parameters $(u_1, u_2)$ are changed to $(-u_1, -u_2)$, the number of degenerate edge-state pairs at quasienergies $0$ and $\pi/T$ are exchanged. This feature facilitates a pairwise analysis of the topological phase patterns.

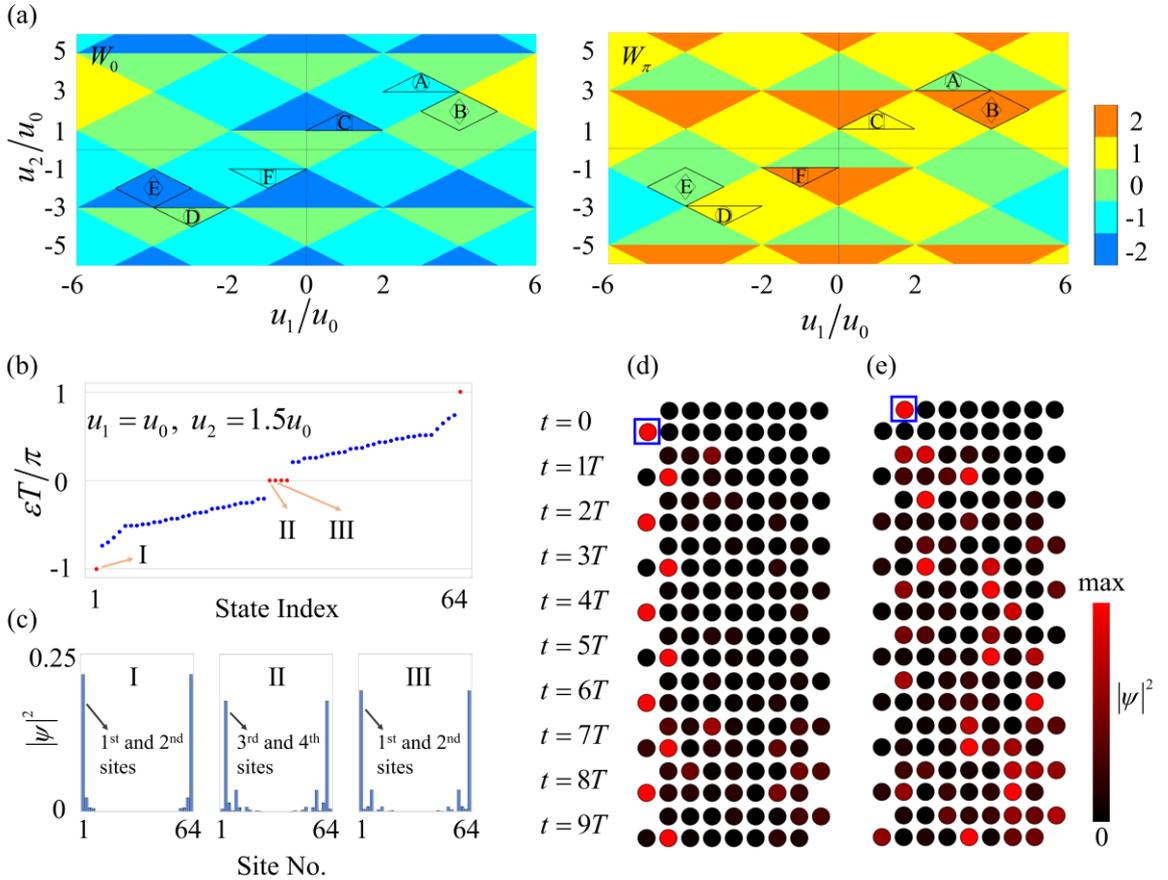

**FIG. 2.** Winding number distributions and results for a representative parameter set. (a) Distribution of the winding numbers $W_0$ (left panel) and $W_\pi$ (right panel) as functions of the intercell coupling parameters $u_1$ and $u_2$. Different winding numbers are indicated by different colors. We label six distinct phases using letters and geometric symbols. (b) Quasienergy spectrum for $u_1 = u_0$ and $u_2 = 1.5u_0$. The topological edge modes are plotted in red. (c) Eigenmode profiles of the three edge states identified in (b). (d)(e) Time-ordered evolution for different initial excitation sites (marked by blue boxes) under the tight-binding approximation. The parameters are set the same as in (b). The size of the finite system is $N = 32$ for (b)(c) and $N = 8$ for (d)(e).

When the NNN coupling is absent, i.e., $u_2 = 0$, the winding number pairs $(0,\ 0)$ and $(-1,\ 1)$ alternate periodically as $u_1$ varies. It indicates that the conventional Floquet SSH model admits only a single topological phase configuration $(-1,\ 1)$. When only the NNN coupling is present, i.e., $u_1 = 0$, the model similarly supports only a single topological configuration, namely $(-2,\ 2)$. For the static extended SSH chain with both NN and NNN couplings, only two topological phases are supported. They are corresponding to winding numbers of 1 or 2 for the zero-energy gap [46]. The introduction of either periodic driving or NNN coupling alone gives rise to only one new topological phase. However, we observe that the presence of a $\pi/T$ quasienergy gap adds an extra degree of freedom to the characterization of topological phases. And the introduction of NNN coupling leads to at most two windings around the origin in the complex plane. Concerning with the absolute values of the winding numbers only, the model with both periodic driving and NNN coupling is expected to exhibit up to $3\times3=9$ distinct phases. Excluding the previously mentioned configurations $(0,\ 0)$, $(-1,\ 1)$ and $(-2,\ 2)$, we also identify six additional topological phases in Fig. 2(a). Specifically, we label each topological phase with a letter and a geometric marker: A

$(-1, 0)$ and D $(0, 1)$ are marked with circles, B $(0, 2)$ and E $(-2, 0)$ with diamonds, and C $(-2, 1)$ and F $(-1, 2)$ with squares. Under OBC, these phases exhibit unequal numbers of degenerate edge-state pairs at quasienergies 0 and $\pi/T$. Their dynamical evolution properties show minor differences compared to phases with equal degenerate edge-state pairs, as will be briefly discussed in Sec. IV.

We show the quasienergy spectrum and the corresponding edge-state distributions for a representative parameter set $u_1 = u_0$, $u_2 = 1.5u_0$ [see Figs. 2(b) and 2(c)]. From the quasienergy spectrum, one can observe a pair of degenerate edge states at $\pi/T$ quasienergy and two pairs at zero quasienergy. This is consistent with the results predicted by the topological numbers $(-2, 1)$. We assign index numbers to the three edge-state pairs. As we can see, the I and III topological edge states are localized at the first and second sites on both ends of the chain, while the II is localized at the third and fourth sites. These two pairs degenerate edge states at zero quasienergy exhibit different boundary localization patterns.

### III. ANOMALOUS EDGE STATES

The quasienergy spectra obtained by fixing $u_2 = 0.8u_0$ and varying $u_1$ within the range $[0, 6u_0]$ are shown in Fig. 3(a). The model undergoes six topological phase transitions. From left to right, the number of degenerate edge-state pairs at zero and $\pi/T$ quasienergies, denoted as $(n_0, n_\pi)$, sequentially takes the values: $(0, 0)$, $(0, 1)$, $(1, 1)$, $(0, 1)$, $(0, 0)$, $(0, 1)$, $(1, 1)$. This result shows perfect agreement with the predictions based on the absolute values of the winding numbers. Interestingly, within the region $u_1 \in [2.2u_0, 5.8u_0]$ where $W_0 = 0$, we observe edge states separating from the bulk bands. These states possess nonzero quasienergies and cross

$\pm\pi/3T$. Similarly, the quasienergy spectra with fixed $u_2 = -0.8u_0$ is presented in Fig. 3(b). The number of degenerate edge-state pairs at zero and $\pi/T$ quasienergies follows the sequence: $(0, 0)$, $(1, 0)$, $(1, 1)$, $(1, 0)$, $(0, 0)$, $(1, 0)$, $(1, 1)$. This result is consistent with the symmetry relation of the phase diagram analyzed in the previous section. In the region $u_1 \in [2.2u_0, 5.8u_0]$ where $W_\pi = 0$, anomalous edge states also separate from the bulk bands. The absolute values of their quasienergies are smaller than $\pi/T$ and traverse $2\pi/3T$. We further fix $u_1 = 3u_0$ and plot the quasienergy spectra over the parameter range $u_2 \in [-2.5u_0, 2.5u_0]$. As shown in Fig. 3(c), the system undergoes three topological phase transitions. The numbers of degenerate edge-state pairs are $(2, 1)$, $(1, 0)$, $(0, 1)$ and $(1, 2)$ from left to right. We can see that the values of $n_0$ and $n_\pi$ exhibit symmetry with respect to $u_2 = 0$. The anomalous edge states near zero and $\pi/T$ quasienergy symmetrically separate from the bulk. But they are not captured by the winding numbers defined at zero and $\pi/T$ quasienergies. These anomalous edge states include modes with quasienergies $\pm\pi/3T$ and $\pm2\pi/3T$, which we refer to as the $\pm\pi/3$ and $\pm2\pi/3$ modes, respectively. The emergence of these two types of edge modes is expected to induce three-period dynamics. We will explore this behavior in Sec. IV.

The mechanism underlying the emergence of the $\pm\pi/3$ and $\pm2\pi/3$ modes differs from that of the recently studied $\pm\pi/2$ modes. The $\pm\pi/2$ modes were discovered in the square-root Floquet SSH model. It is a four-band system featuring two additional gaps at quasienergies $\pi/2T$ and $-\pi/2T$ [41]. These two additional quasienergy gaps allow for the existence of topological edge modes. And these modes can be characterized by the topological invariant of the parent model associated with the zero gap [41-43]. However, the $\pm\pi/3$ and $\pm2\pi/3$ edge modes are merely two special types among the anomalous edge states we have identified. In our model, no continuous parameter region is found to support the persistent existence of either $\pm\pi/3$ or $\pm 2\pi/3$ modes. Similar edge modes have not been reported in one-dimensional SSH models that incorporate LRCs or periodic driving individually. We speculate that these anomalous edge states originate from certain effective defects induced by the combined presence of the NNN coupling and periodic driving.

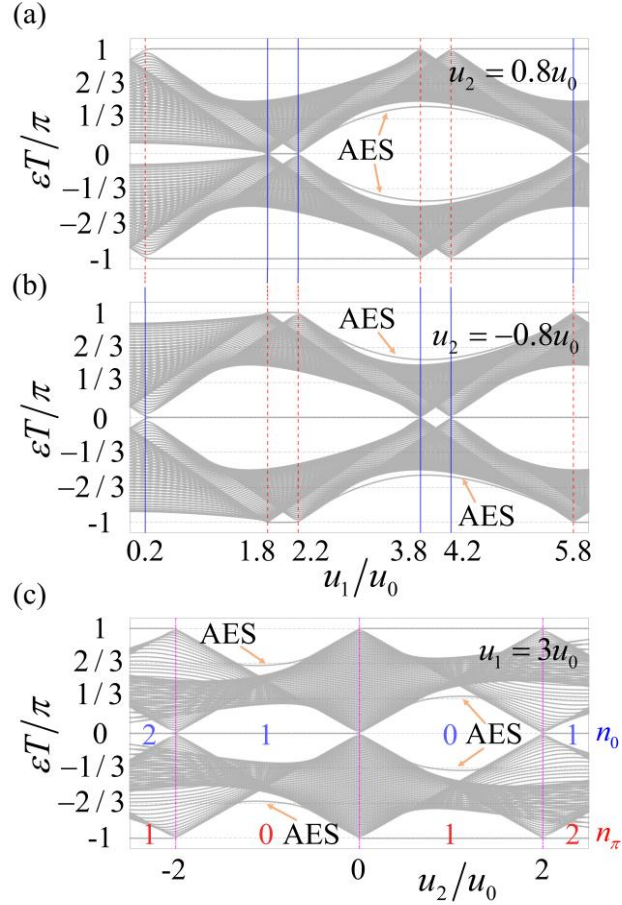

**FIG. 3.** Quasienergy spectra with anomalous edge states (AES). (a)(b) Quasienergy spectra as functions of $u_1$ with fixed parameters $u_2 = 0.8u_0$ and $u_2 = -0.8u_0$, respectively. Blue solid (red dashed) lines indicate the phase boundaries of the winding number $W_0$ ($W_\pi$). (c) Quasienergy spectra as a function of $u_2$ with fixed $u_1 = 3u_0$. The purple dotted lines mark simultaneous topological phase transitions of $W_0$ and $W_\pi$. All results are obtained for a system size of $N = 64$.

## IV. PERIODIC EVOLUTION FEATURE

### A. Evolutions under the tight-binding approximation and analyses to the periodic evolution feature

The evolution of nonequilibrium quantum systems can be performed using the Dyson time-evolution formula [54]

$$\psi(t) = \mathrm{T}\,(e^{-i\int_0^t H(\tau)d\tau})\psi(0). \tag{12}$$

Subsequently, we use this formula to perform evolution under the tight-binding approximation in order to observe the two-period and three-period dynamics of the Floquet extended SSH system. The initial state is chosen as a localized excitation at site $j_0$, represented by the basis vector $\psi(0) = |j_0\rangle$. This corresponds to a quantum state with unit amplitude at site $j_0$ and zero elsewhere.

We begin by discussing the results of the two-period evolution. The quasienergy spectrum and eigenmode profiles for the parameter setting $u_1 = u_0$, $u_2 = 1.5 u_0$ have been analyzed previously. Here, we investigate the corresponding dynamical properties. As shown in Fig. 2(d), when an initial excitation is applied at the first site on the left end of the chain, a clear boundary localization is observed. It is accompanied by a two-period oscillation between the first and second sites. We can understand the $2T$ periodicity as follows. The initial excitation is a superposition of the I ($\pi$) and III (zero) edge modes, denoted as $\psi(0) = |\psi\rangle = c_1|0^{\mathrm{III}}\rangle + c_2|\pi^{\mathrm{I}}\rangle$. After two driving periods

$$\begin{aligned} U_T|\psi\rangle &= c_1|0^{\mathrm{III}}\rangle - c_2|\pi^{\mathrm{I}}\rangle, \\ U_{2T}|\psi\rangle &= c_1|0^{\mathrm{III}}\rangle + c_2|\pi^{\mathrm{I}}\rangle, \end{aligned} \tag{13}$$

the original profile can be restored. In contrast, when the initial excitation is applied at the site where the edge state II is localized, the excitation rapidly spreads into the bulk without period-doubled evolution [see Fig. 2(e)]. The zero edge mode II is unpaired with any $\pi$ edge mode. It indicates that the periodic evolution is not always guaranteed for our Floquet system with multiple pairs of edge states. It requires the initial excitation to be placed at a site where edge modes with different quasienergies are simultaneously localized.

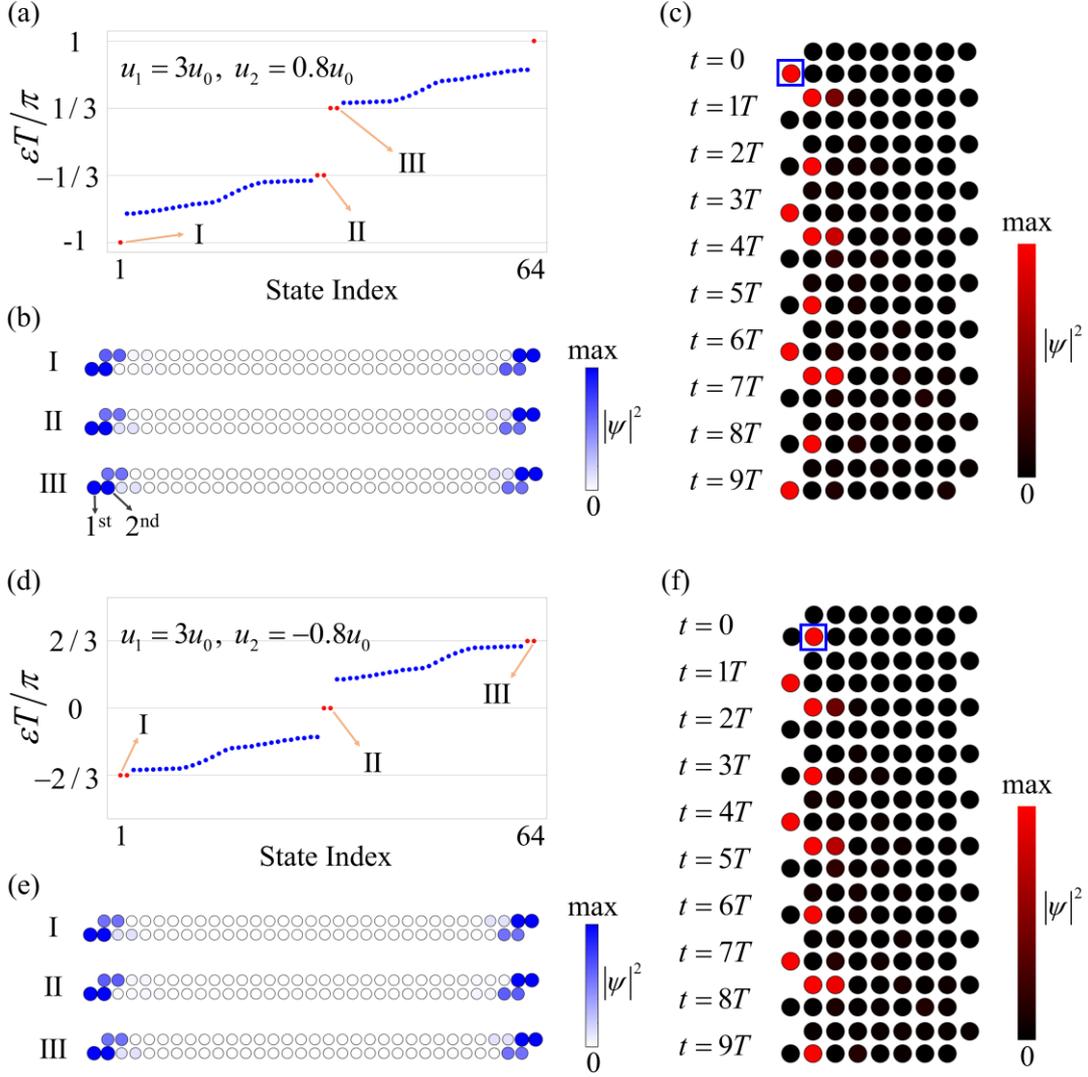

FIG. 4. The three-period evolutions under the tight-binding approximation. (a)(d) Quasienergy spectra for two parameter sets $u_1 = 3u_0$, $u_2 = 0.8u_0$ and $u_1 = 3u_0$, $u_2 = -0.8u_0$. These exhibit $\pm\pi/3$ and $\pi$ modes, $\pm 2\pi/3$ and zero modes, respectively. (b)(e) Corresponding eigenmode profiles of (a) and (d). The size of the finite system is $N = 32$. (c) The $3T$-periodic evolution with $\pm\pi/3$ and $\pi$ modes. (f) The $3T$-periodic evolution with $\pm 2\pi/3$ and zero modes. The initial excitation site is marked with a blue box, and the size of the system in the evolution is set to $N = 8$.

Now, we analyze the dynamics of three-period evolution. At the parameter setting $u_1 = 3u_0$ and $u_2 = 0.8u_0$, the system hosts $\pm\pi/3$ and $\pi$ modes. Their eigenmode profiles are all localized at the first and second sites on both ends of the chain [see Figs. 4(a) and 4(b)]. We apply an initial excitation at the first site and perform numerical evolution. As shown in Fig. 4(c), the excitation returns to the initial site every three driving periods. The three-period evolution observed here can be understood through the following expression:

$$\begin{aligned}
\psi(0) &= |\psi\rangle = c_1|\pi\rangle + c_2|-\pi/3\rangle + c_3|\pi/3\rangle, \\
U_T|\psi\rangle &= e^{-i\pi} \cdot c_1|\pi\rangle + e^{i\frac{1}{3}\pi} \cdot c_2|-\pi/3\rangle + e^{-i\frac{1}{3}\pi} \cdot c_3|\pi/3\rangle, \\
U_{2T}|\psi\rangle &= e^{-i2\pi} \cdot c_1|\pi\rangle + e^{i\frac{2}{3}\pi} \cdot c_2|-\pi/3\rangle + e^{-i\frac{2}{3}\pi} \cdot c_3|\pi/3\rangle, \\
U_{3T}|\psi\rangle &= -(c_1|\pi\rangle + c_2|-\pi/3\rangle + c_3|\pi/3\rangle).
\end{aligned} \qquad (14)$$

Symmetrically, as shown in Figs. 4(d) and 4(e), the system hosts $\pm 2\pi/3$ and zero modes under the parameter setting $u_1 = 3u_0$ and $u_2 = -0.8u_0$. Their eigenmode profiles are likewise localized at the first and second sites on both ends of the chain. We apply the initial excitation at the second site. According to expression

$$\begin{aligned}
\psi(0) &= |\psi\rangle = c_1|0\rangle + c_2|-2\pi/3\rangle + c_3|2\pi/3\rangle, \\
U_{3T}|\psi\rangle &= c_1|0\rangle + e^{i2\pi} \cdot c_2|-2\pi/3\rangle + e^{-i2\pi} \cdot c_3|2\pi/3\rangle \\
&= c_1|0\rangle + c_2|-2\pi/3\rangle + c_3|2\pi/3\rangle,
\end{aligned} \qquad (15)$$

the numerical evolution exhibits a three-period behavior [see Fig. 4(f)]. In fact, such behavior can be observed when the excitation is applied at either the first or second site. Here, we present only one case for each parameter configuration.

The two-period and three-period evolution observed here merely mean that the intensity maximum returns to the initial excitation site after specific multiples of the driving period. They are not perfectly periodic oscillations with uniform amplitude. This is because the interfering edge modes are not identical. They are simply localized at the same boundary sites, rather than sharing exactly the same eigenmode profiles. To more clearly reveal the periodic localization of the intensity maximum, we present results of a longer-time evolution in the Appendix.

## B. The three-period evolution in photonic waveguides simulation

Next, we investigate the three-period evolution in photonic lattice simulation. By rearranging the model shown in Fig. 1, we construct the photonic lattice illustrated in Fig. 5(a). The time-dependent couplings are implemented by modulating the spacing between adjacent waveguides along the propagation direction $z$. A Schrödinger-like equation

$$i\partial z\psi = -\frac{1}{2k_0}\nabla_\perp^2\psi - \frac{k_0\delta n(x,y,z)}{n_0}\psi \qquad (16)$$

can be used to described the photonic lattice. In this context, $\nabla_\perp^2 = \partial^2 x + \partial^2 y$, $k_0 = 2\pi n_0/\lambda$ denotes the wave number, $\psi$ represents the field distribution, $n_0$ is the background refractive index and $\delta n$ stands for the refractive index detuning. The refractive index modulation near the waveguide is described by

$$\delta n(x,y,z_0) = A e^{-[(\delta x/\sigma)^2+(\delta y/\sigma)^2]^3}, \qquad (17)$$

where $\sigma$ is radius and $(\delta x, \delta y)$ is the center coordinates of the waveguides in the $xy$ plane at propagation distance $z_0$. In the following simulation, the parameters are set to $n_0 = 2.1$, $\lambda = 1550\,nm$, $A = 2.6\times10^{-3}$ and $\sigma = 4.9\,\mu m$. Additionally, the waveguide length of one period is set to $L = 40\,mm$.

The wave equation is solved using the split-step Fourier method [55]. Following the simulation method provided in Ref. [43], we design optical simulation of the three-period evolution in the Floquet extended SSH model. The designed separation of two waveguides in one period is given by

$$d = \begin{cases} a & 0 < z \leq 0.5L \\ \begin{cases} a_x - \dfrac{1+\cos\phi}{2}(a_x - d_{cx}) \\ a_y - \dfrac{1+\cos\phi}{2}(a_y - d_{cy}) \end{cases} & 0.5L < z \leq 0.6L \\ d_c & 0.6L < z \leq 0.9L \\ \begin{cases} a_x - \dfrac{1-\cos\phi}{2}(a_x - d_{cx}) \\ a_y - \dfrac{1-\cos\phi}{2}(a_y - d_{cy}) \end{cases} & 0.9L < z \leq L \end{cases} \quad (18)$$

Here, $\phi = 10\pi z/L$, $a = \sqrt{a_x^2 + a_y^2}$ is the initial waveguide separation ensuring negligible coupling, and $d_c = \sqrt{d_{cx}^2 + d_{cy}^2}$ denotes the tunable separation between two waveguides to realize to a targeted coupling parameter. In Step2, the coexistence of two types of intercell couplings results in a relatively compact arrangement of the waveguide array. If the waveguide spacings $d_c$ associated with the NN and NNN couplings are too small, their sum may fall below the threshold distance $a$. In this case, additional unwanted couplings will be introduced to the photonic waveguide system, causing the system to fail to realize the intended model. Therefore, we refer to the parameters in Ref. [43], and adjust the background refractive index $n_0$ and the waveguide length within one period $L$, as mentioned above. Under those settings, the coupling strength decreases more rapidly with increasing waveguide spacing. As a result, the coupling between waveguides can be regarded as negligible when $a = 32\ \mu m$. Moreover, we have $d_c = 17.54\ \mu m$ corresponds to $u_0 = \pi/2$, $d_c = 13.80\ \mu m$ corresponds to $u_1 = 3u_0$, and $d_c = 18.64\ \mu m$ corresponds to $u_2 = 0.8u_0$. These spacing parameters could ensure that both NN and NNN couplings are realized between photonic waveguides without introducing any additional unwanted couplings during Step2.

With all parameters obtained, and based on Eq. (17), we plot the refractive index distributions of the waveguides at different steps shown in Fig. 5(b). From the uncoupled state of two waveguides ($0 < z \leq 0.5L$) to either Step1 or Step2, the movement direction of the waveguides is not along the $x$ axis or $y$ axis. Therefore, our waveguide bending scheme differs slightly from that in the reference in the transition stages $0.5L < z \leq 0.6L$ and $0.9L < z \leq L$ [see Eq. (18)]. In addition, each waveguide located inside the lattice needs to couple with two other waveguides simultaneously in Step2. It is difficult to find a trajectory that both the NN and NNN couplings simultaneously satisfy Eq. (18) during the transition stage. In the simulation, we ensure the NN couplings strictly follows equation (18). As for the NNN couplings, we only ensure that the waveguide spacing meets the parameter requirement during the stable coupling stage ($0.6L < z \leq 0.9L$).

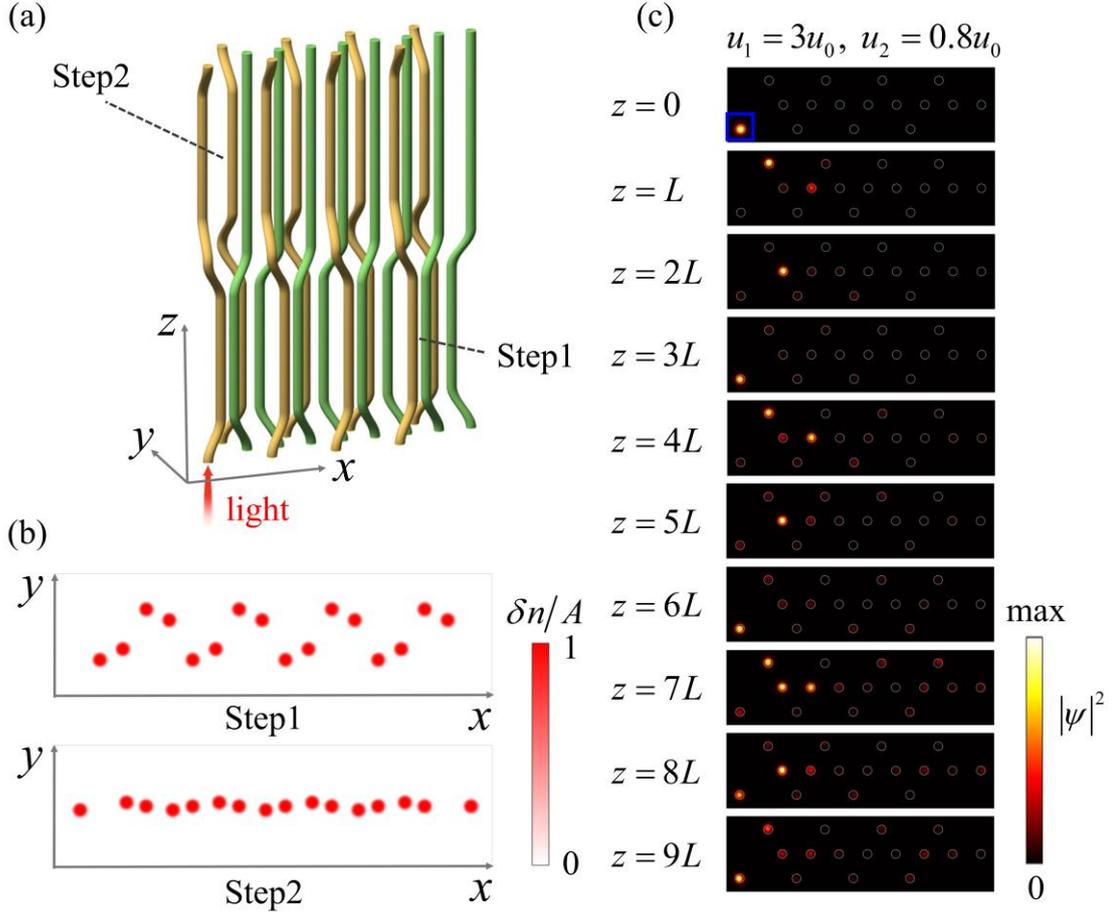

**FIG. 5.** Evolution in a photonic Floquet extended SSH waveguide array. The size of the system is $N = 8$. (a) Schematic sketch of the photonic waveguide array. The initial light is injected into the first waveguide. (b) Refractive index distributions of waveguides in Step1 and Step2. (c) The $3L$-

periodic evolution with $\pm\pi/3$ and $\pi$ modes.

However, most of the field propagation occurs during the stable coupling stage. In a short-time simulation, the slight imperfections can be neglected. We apply an initial excitation at the first site (marked by blue box) and observe its evolution over nine driving periods. As illustrated in Fig. 5(c), the light periodically localizes at the first waveguide every three period, which shows a three-period evolution. With carefully adjusted optical parameters, we successfully realize a Floquet SSH model with NNN coupling in a photonic waveguide array and observe three-period evolution in the simulation for the first time.

## V. CONCLUSION

To conclude, we construct a Floquet extended SSH model, where anomalous edge states beyond the description of winding numbers were identified. These include edge modes with quasienergies $\pm\pi/3T$ and $\pm 2\pi/3T$. Through the incorporation of carefully adjusted optical parameters, we simulate a photonic Floquet waveguide system with NNN coupling, in which the characteristic three-period evolution was successfully observed. The $\pm\pi/3$ and $\pm 2\pi/3$ modes discussed here are anomalous edge states rather than topological ones, and they are distinct from the recently studied topological $\pm\pi/2$ modes [39-43]. Nevertheless, they also lead to corresponding period-multiplied evolution. In addition, our model exhibits topological phases with unequal numbers of zero and $\pi$ edge modes. The observability of period-doubled dynamics is related to whether the initial state is a superposition of multiple eigenmodes.

Our work may offer a new perspective for constructing models that exhibit more diverse period-multiplied dynamics. Researchers can induce anomalous edge states with specific quasienergies by breaking certain symmetries of the model or introducing structural defects. It may be more feasible than creating multiple energy gaps that allow for topological characterization. Moreover, our work can offer a degree of experimental relevance and potential applicability. The simulation design is aligned with realistic physical systems, and all necessary parameters have been specified. In light of the notable progress in implementing topological systems via laser-written waveguide arrays [15, 17, 18, 20, 21, 56], our results could be accessible to experimental verification. If such anomalous edge states can eventually be realized in quantum lattice, they may also find promising applications in quantum computing [57, 58].

## ACKNOWLEDGMENTS

The authors thank for the support by National Natural Science Foundation of China under (Grant 12404365).

# APPENDIX: LONGE TIME EVOLUTION ANALYSIS

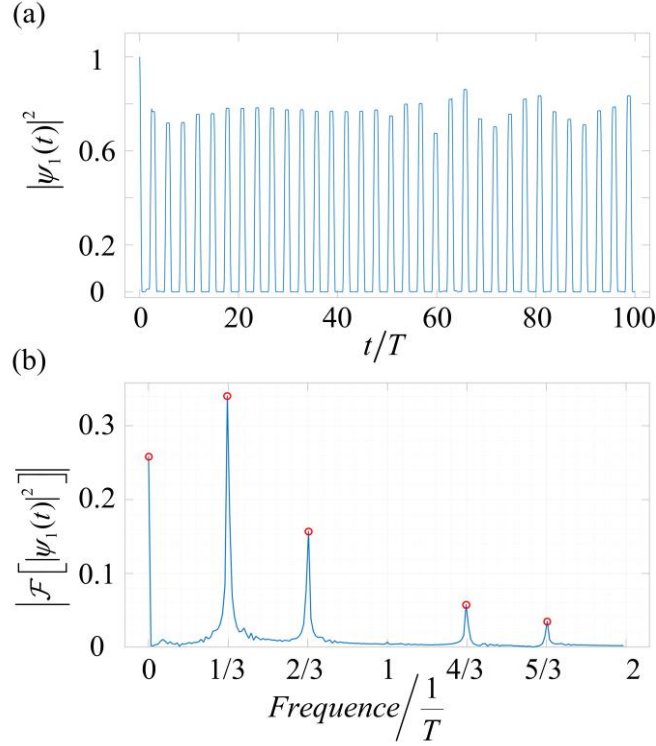

**FIG. S1.** A longer-time evolution under the tight-binding approximation. The initial excitation is put at the first site, the size of the system is $N = 32$, and the parameters are $u_1 = 3u_0$, $u_2 = 0.8u_0$. (a) The result of evolution over $100T$. The vertical axis represents the intensity at the first site over time. (b) Fourier transform of the data in (a).

We present evolution results under the tight-binding approximation over $100T$. The parameters are set to $u_1 = 3u_0$ and $u_2 = 0.8u_0$, which exhibit three-period dynamics. In a finite system with $N = 32$ sites, we apply an initial excitation at the first site and record the intensity at that site over time. As shown in Fig. S1(a), the intensity exhibits a square-wave-like temporal profile, reaching peak values at regular time intervals. All peak values exceed 0.6 and it indicates a characteristic of periodic localization. However, it is also evident that these peak values are not identical. So, it is not strictly periodic in the rigorous sense.

In addition, we perform a Fourier transform on the time-domain data and obtain the corresponding frequency spectrum as shown in Fig. S1(b). Excluding the zero-frequency (DC) component, we observe relatively high values at integer multiples of 1/3 in the frequency spectrum. It indicates that, in the time domain, the peak values appear once every three driving periods. Nevertheless, the spectrum is continuous and also contains various other frequency components. It also implies that the time-domain data of intensity is not strictly periodic with constant amplitude. Combining the time-domain and frequency-domain results, our understanding of the periodicity merely means that the intensity maximum returns to the initial excitation site after specific multiples of the driving period.